**On the problem of phase transitions in lysozyme crystals**

Vasylkiv Yu., Nastishin Yu. and Vlokh R.

Institute of Physical Optics, 23 Dragomanov St., 79005 Lviv, Ukraine
e-mail: vlokh@ifo.lviv.ua

**Abstract**

We present experimental evidence of the fact that lysozyme crystals, which are grown from their mother solution and exist in it, dissolve on heating above $T \approx 307\,\text{K}$. We argue that the anomaly in the light scattering recently observed at the temperature $T = 307\,\text{K}$ and identified in the reference [Svanidze A. V. *et al.* 2006. JETP Lett. **84**: 551] as a structural crystalline phase transition in the single lysozyme crystals, in fact, corresponds to a temperature limit of the crystal existence.

**Introduction**

Recently we have shown experimentally [1,2] that hen lysozyme crystals grown from their mother solution in a glass capillary exhibit a special temperature point $T_c = 284\,\text{K}$, where the interference colours of the crystal texture observed with the polarization microscopic technique change drastically, implying a step-like change in the optical retardation of this crystal. The textural changes are reversible when cycling the temperature and they have been explained as a phase transition. We have also found that the polarization microscopic textural changes alluded to above are accompanied by a significant increase in the solution turbidity. It is now well established (see, e. g., [3]) that such an increase in the lysozyme solution turbidity is associated with self-aggregation of lysozyme molecules. The temperature at which the turbidity drastically increases is usually termed in the literature as a "cloud point".

On the other hand, recently an observation of structural phase transition in the lysozyme crystals has been reported at $T = 307\,\text{K}$ [4] on the basis of Brillouin scattering technique. The authors of the work [4] have registered anomalous temperature dependence of quasi-transverse phonon velocity in the vicinity of this temperature. As an additional argument supporting their conclusion on the existence of phase transition at $T = 307\,\text{K}$, the authors [4] have referred to the study [5], where a jump-like change in the optical phase retardation has earlier been observed at $T = 306.5\,\text{K}$. Here it is worth noticing that the authors of the mentioned work [5] have stated that the specimen (but not the lysozyme molecules) "began to be denatured above this temperature". Let us take into account that the solubility of lysozyme increases on heating. Then it would be natural to accept that the observed size decrease for the crystal samples of lysozyme is due to dissolving of the lysozyme molecules. The aim of this paper is to present the results evidencing that the temperature $T = 307\,\text{K}$ corresponds to beginning of dissolution of the lysozyme crystal in the mother solution rather than a true structural phase transition between some crystalline modifications, as claimed in [4].

**Results and discussion**

Fig. 1 represents microphotographs of the lysozyme crystals, which are grown from their mother solution and coexist with it in a glass capillary. The growth conditions are the same as described in our previous works [1,2]. The photographs in the sequence adopted in Fig. 1, from left to right, have been taken on heating at different temperatures indicated in the figure caption. For all the photographs, the specimens remain in the environment of their mother solution. It is seen from Fig. 1a–c that in the temperature region $T = 293-306\,\text{K}$ the lysozyme crystals exhibit a specific habit, with well developed growth terraces on the crystal surface. However, the surfaces of the specimens become more and more erosive, starting approximately from $308\,\text{K}$ (see Fig. 1d), and the crystal size apparently decreases, indicating dissolving of the crystal specimens. The samples become significantly smaller in size upon further heating in the temperature region $311-315\,\text{K}$ (see Fig. 1e–g). They scatter light progressively and finally disappear, being dissolved in the mother solution at $323\,\text{K}$ (Fig. 1i) and leaving a diffusive spot after them.

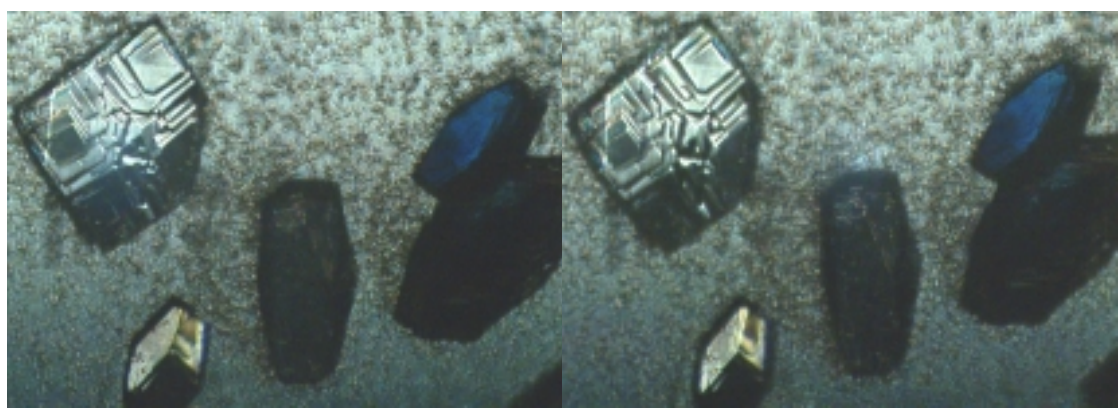

a　　　　　　　　　　　　　b

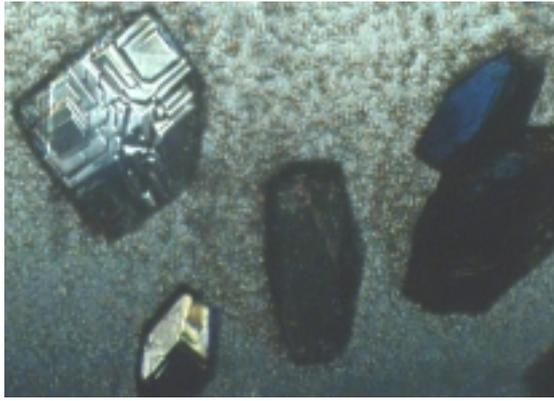 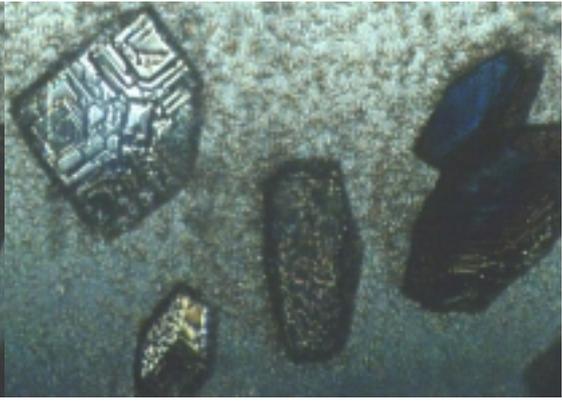

c                           d

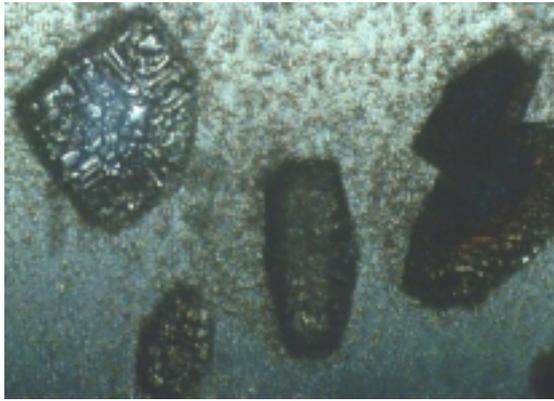 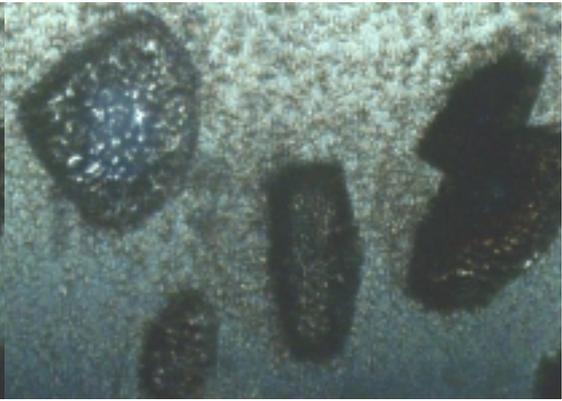

e                           f

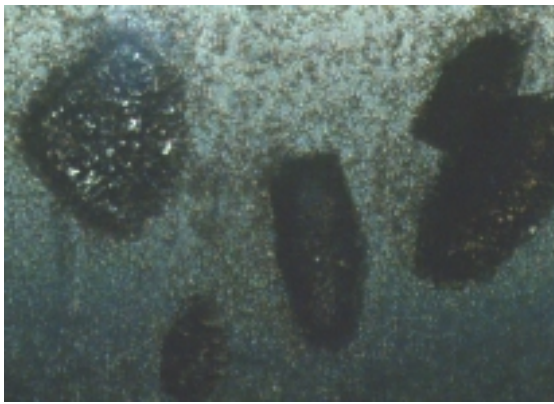 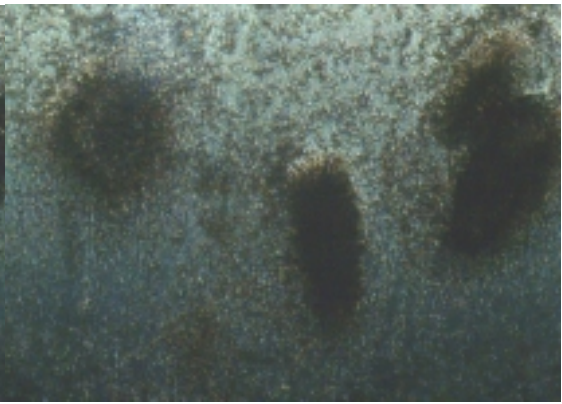

g                           h

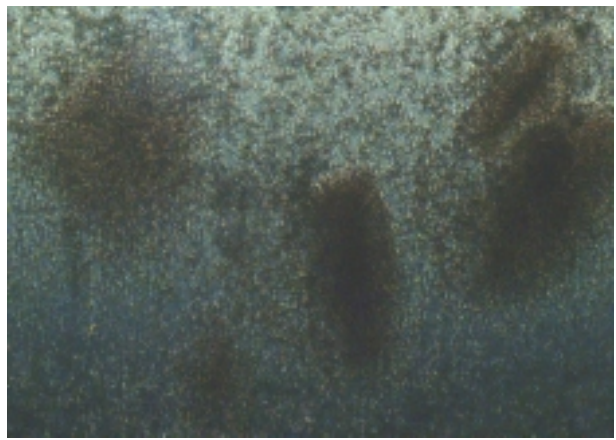

Figure 1. Microphotographs of lysozyme crystals taken at different temperatures: a – $T = 293\,\text{K}$, b – $298\,\text{K}$, c – $306\,\text{K}$, d – $308\,\text{K}$, e – $311\,\text{K}$, f – $313\,\text{K}$, g – $315\,\text{K}$, h – $317\,\text{K}$ and i – $323\,\text{K}$.

We thus are led to conclude that the lysozyme crystals begin to dissolve in the mother solution at the temperature $T = 307\,\text{K}$ and, hence, no crystalline structural phase transition could take place at this temperature point. Notice that erosion of the crystal surfaces accompanying the dissolving process can be a reason for the anomalous behaviour of the Brillouin scattering observed in [4], as well as a reason for the lack of any birefringence data above $T = 307\,\text{K}$ in the reference [5].

Concerning the phase transitions in the lysozyme crystals, two groups of references may be quoted at present, namely [1,2] and [6,7]. Our observations performed with the aid of polarization microscopy [1,2] have indicated textural changes implying a phase transition in the lysozyme crystal, accompanied by increased turbidity in the mother solution, and indicating the "cloud point" at $T = 284\,\text{K}$. The authors of Ref. [3] have characterized the mother solution by differential scanning calorimetry and explained the "cloud point" by self-aggregation of the lysozyme molecules. The aggregation of the latter in the mother solution occurred in the vicinity of the "cloud point" can be related to the textural changes observed in the lysozyme crystal, which coexists with the mother solution at the same temperatures. In other words, these two phenomena could be driven by the same mechanism. The beginning of the aggregation of the molecules in the mother solution implies new interactions that bind the molecules together. The very same intermolecular interaction mechanism can be responsible for the structural transformation in the crystal observed at the same temperature.

In Refs. [6,7] it is reported that the lysozyme crystals grow in two modifications. Namely, below $298\,\text{K}$ one finds a tetragonal form ($P4_32_12$) and above this temperature single crystals grow in an orthorhombic form ($P2_12_12_1$). The authors [6,7] claim that "on heating at $298\,\text{K}$ the tetragonal form transforms into the orthorhombic". In Ref. [8] these authors suppose that the tetragonal form is in a metastable state. At the same time, according to the X-ray data [7] for a dry tetragonal lysozyme crystal, removed from the mother solution and kept at $313\,\text{K}$, the tetragonal form remains stable for several days and shows no transformations. According to [6,7], the tetragonal-to-orthorhombic transformation passes through a dissolved state and so it is hard to qualify such a transformation as a crystalline structural phase transition. In compliance with the X-ray data [6,7], no other phase transitions above the room temperature have been documented for the both forms of the lysozyme crystals. Taking into account our results presented above, it remains to accept that the anomaly in the light scattering reported in [4] at $T = 307\,\text{K}$ corresponds to the temperature-induced dissolving of the crystal, which is illustrated in Fig. 1 and is observed at the very same temperatures.

**Conclusions**

We have demonstrated that the lysozyme crystals, which have been grown in their mother solution and then exist in it, disappear on heating starting from $307\,\text{K}$, via the temperature-induced dissolving in the mother solution. This observation could be directly related to the anomaly in the light scattering detected in Ref. [4] and claimed there as a crystalline phase transition between two different crystal modifications. Our opinion is that the temperature point $T = 307\,\text{K}$ has been misidentified in [4] as a temperature of crystalline phase transition, whereas in fact this peculiar point might be caused by the temperature-induced dissolution of the crystals.